\documentclass[11pt]{article}
\usepackage{setspace}
\usepackage{authblk} 
\usepackage[utf8]{inputenc} 
\usepackage[T1]{fontenc}    
\usepackage{amsfonts}       
\usepackage{amssymb}
\usepackage{amsmath}
\usepackage{booktabs}       
\usepackage{hyperref}       
\usepackage{url}            
\usepackage{microtype}      
\usepackage{nicefrac}       
\usepackage{paralist}       
\usepackage{graphicx}
\usepackage{float}
\usepackage{wrapfig}
\usepackage[dvipsnames]{xcolor}
\usepackage{ulem}
\usepackage{tikz}
\usetikzlibrary{arrows,shapes,angles,quotes}
\usetikzlibrary{decorations.pathreplacing}
\usepackage{fullpage}
\usepackage{lmodern}
\usepackage{ulem}

\usepackage{subcaption}

\newcommand*{\vcenteredhbox}[1]{\begin{tabular}{@{}l@{}}#1\end{tabular}}
\begin{document}
\title{Homogeneous Nucleation of Undercooled Al-Ni melts via a Machine-Learned Interaction Potential}
\author[1,2,4]{Johannes Sandberg}
\author[1,2]{Thomas Voigtmann}
\author[3]{Emilie Devijver}
\author[4]{Noel Jakse}
\affil[1]{Institut für Materialphysik im Weltraum, Deutsches Zentrum für Luft- und Raumfahrt (DLR), 51170 	 Köln, Germany }
\affil[2]{Department of Physics, Heinrich-Heine-Universität Düsseldorf, Universitätsstraße 1, 40225 Düsseldorf, Germany}
\affil[3]{CNRS, Univ. Grenoble Alpes, Grenoble INP, LIG, F-38000 Grenoble, France}
\affil[4]{Université Grenoble Alpes, CNRS, Grenoble INP, SIMaP F-38000 Grenoble, France}
\date{}

\maketitle

\begin{abstract}
    Homogeneous nucleation processes are important for  understanding  solidification and the resulting microstructure of materials.
    Simulating this process requires accurately describing the  interactions between atoms, which is further complicated by chemical order through cross-species interactions. 
    The large scales needed to observe rare nucleation events are far beyond the capabilities of \textit{ab initio} simulations.
    Machine-learning is used for overcoming these limitations in terms of both accuracy and speed, by building a high-dimensional neural network potential for binary Al-Ni alloys, which serve  as a model system relevant to many industrial applications.
    The potential is validated against experimental diffusion, viscosity, and scattering data, and is  applied to large-scale molecular dynamics simulations of homogeneous nucleation at equiatomic composition, as well as for pure Ni. 
    Pure Ni nucleates in a single-step into an fcc crystal phase, in contrast to previous results obtained with a classical empirical potential. This highlights the sensitivity of nucleation pathways to the underlying atomic interactions.
    Our findings suggest that the nucleation pathway for AlNi proceeds in a single step toward a B2 structure, which is discussed in relation to the pure elements counterparts. 
    
\end{abstract}

\section{Introduction}

The macroscopic properties of a material are closely tied to its microstructure, and understanding their relationship is a central topic in materials science \cite{mittemeijer2010fundamentals}. 
On a fundamental level, the observed properties originate from the chemical and physical interactions that shape the atomic arrangements and ultimately determine these properties. 
It is therefore vital to understand the microscopic nature of crystal nucleation, and growth when the liquid morphs into the solid, if one wants to control the microstructure.
Especially, the early steps of the microscopic nucleation process cannot generally be directly observed in experiment, with the very recent exception for Fe-Pt binary metallic nanoparticles by atomic electron tomography \cite{zhou2019observing} that still require to be combined with atomic scale simulations. 
Often, systems such as colloids \cite{zhang2014experimental} are use as model systems studied experimentally for nucleation, still necessitating a computational treatment \cite{sosso2016crystal}. Due to the presence of solvent-mediated many-body interactions, it is also not clear whether the kinetic pathways are the same in colloids as in metallic or molecular systems \cite{Radu2014}.
Simulations at the atomic scale such as Molecular Dynamics (MD) remain one of the dedicated tools that allow for the direct \textit{in silico} study of critical nuclei, their growth, and the local structures in which they emerge from the melt, as reported for generic models as well as metals and alloys \cite{auer2001prediction,kawasaki2010formation,toxvaerd2020role, tang2013anomalously, shibuta2017heterogeneity, fujinaga2020nucleation, becker2020glass, becker2022unsupervised, orihara2020molecular, shang2023influence}.  

Due to the low probability of observing rare crystal nucleation events, large scale simulations are required. For pure metals and alloys of interest here, empirical or semi-empirical interaction potentials, such as embedded atom models \cite{daw1984embedded,daw1993embedded}, modified embedded atom models \cite{baskes1992modified,lee2003semiempirical}, and reactive force fields \cite{chenoweth2008reaxff}, are  traditional choices for such simulations, owing to their high performance.
There are, however, often large discrepancies between such force fields and experimental results \cite{peng2019chemical, sosso2016crystal}.
A better approximation can in principle be achieved by first-principles approaches.
\textit{Ab initio} simulations based on Density Functional Theory (DFT) \cite{hohenberg1964inhomogeneous,kohn1965self}, through Car-Parrinello  \textit{ab initio} molecular dynamics (AIMD) \cite{car1985unified,hafner2008ab}, provide a more faithful representation of the interatomic interaction. Those ab initio simulations  enable the accurate study of arbitrary mixtures of species in various phases, and have seen success across much of material physics \cite{jakse2003local,alemany2004kohn,alemany2007ab,souto2010ab,jakse2013liquid,del2017ab,pasturel2017atomic,del2020structure,alemany2021static,fahs2023structure,cao2024influence}.
Unfortunately, computational cost and poor scaling with system size put \textit{ab initio} simulation at odds with the large scale required for observing rare nucleation events.

Machine Learning Interatomic Potentials (MLIPs) were introduced to overcome the limitations of both classical potentials, and \textit{ab initio} simulations.
Since the seminal work of Behler and Parrinello \cite{behler-parrinello-2007}, a wide range of methods for constructing data-driven potentials have been proposed.
These range from linear-regression-based models \cite{thompson2015spectral,seko2014sparse,takahashi2018linearized,kandy2023comparing}, kernel-based methods such as Gaussian approximation potentials \cite{bartok2010gaussian,bartok2015g,deringer2021gaussian}, descriptor-based neural network approaches such as the Behler-Parrinello High Dimensional Neural Netowrk Potentials (HDNNP) \cite{behler-parrinello-2007,behler-tutorial-2015,behler2021four}, Deep Potentials \cite{zhang2018deep}, and more recently graph neural network potentials \cite{schutt2018schnet,batzner20223}.
By fitting a machine-learning regression model to \textit{ab initio} potential energy surfaces, it is possible to effectively interpolate them at a computational cost that is orders of magnitude lower.
Molecular dynamics studies of nucleation, which require both accurate potentials capable of representing multiple phases and long, large-scale simulations to capture rare nucleation events, benefit greatly from MLIPs  \cite{goniakowski2022nonclassical,jakse2022machine}.

Al-Ni alloys are known for their mechanical performance, thermal stability, and advanced functional properties, making them versatile in both structural and functional applications. 
The structural ordering in Al-Ni melts has therefore drawn a lot of interest over the past decades, aimed at understanding the structure-dynamics relationship, and in particular the impact of chemical short-range ordering \cite{maret1990structure,jakse2004ab,jakse2005chemical,das2005influence,kuhn2014diffusion,jakse2014dynamic,jakse2015relationship}.
Meanwhile, the challenges of experimentally measuring Al self-diffusion coefficients make MD simulations a vital tool for studying this system, and Al alloys more broadly. Molecular dynamics simulations using the first EAM potential by Mishin \cite{mishin2002embedded} were used to study self and inter diffusion in Al-Ni across various compositions.  These simulations \cite{kuhn2014diffusion} found different diffusivities between the two species, even in Ni-rich compositions, in contrast to the \textit{ab initio} results of \cite{jakse2015relationship}.
Recently, classical simulations using a new version of the EAM potential of Mishin \cite{purja2009development} investigated nucleation into the bcc-based B2 structure at equiatomic composition, and the two-step nucleation of pure Ni into fcc via bcc precursors driven by local icosahedral ordering \cite{orihara2020molecular}.
Nucleation at AlNi and AlNi$_3$ compositions were studied using a novel structure identification technique, showing a two-step process for the Ni-rich composition, where chemical ordering of the final phase appearing before bond-orentational ordering during nucleation onset \cite{becker2022crystal}. 

A reliable description of early stages of crystal nucleation requires an accurate description of the structural and the diffusion in the liquid as well as the crystalline states. 
Improvement of the interatomic interactions in Al-Ni alloy at the \textit{ab initio} accuracy can be achieved in the framework of machine learning, which has proven to be able to tackle nucleation phenomena for pure aluminium \cite{jakse2022machine}.

In the present work, homogeneous crystal nucleation pathways are studied for Al\textsubscript{50}Ni\textsubscript{50} as well as pure Ni. 
For this purpose, a HDNNP \cite{behler2021four} is first trained on  AIMD trajectories for Al-Ni alloy in the the whole composition range including pure Ni and pure Al, in the liquid states above and below the melting point as well as relevant crystalline phases at low temperatures. 
The trained potential is first successfully validated on the local structure as well as self- and inter-diffusion, examining the temperature and composition dependence.
Homogeneous nucleation under deep undercooling is then investigated with these potentials by means of large-scale MD simulations with $135000$ atoms for the two compositions. For the equiatomic composition a single-step solidification process into the B2 phase is observed. 
Such a nucleation pathway is discussed in view of the nucleation of their pure elements counterparts.  
Furthermore, in all the cases nuclei show a relatively irregular, non-spherical shape, different from what is assumed in Classical Nucleation Theory (CNT).

The remaining part of the paper is organized as follows. Section \ref{sec:methods} describes the construction of the MLIP and the various properties of interest in the present work. Section \ref{sec:results} is devoted to the validation of the potential as well as the analysis of  homogeneous nucleation for the two compositions as well as pure Ni. The conclusions are provided in Section \ref{sec:conclusion}.

\section{Computational background}
\label{sec:methods}
\subsection{Dataset: \textit{Ab initio} molecular dynamics trajectories}
Care must be taken in designing a dataset from \textit{ab initio} calculations, to ensure that the configurations added to the dataset are representative of those encountered during a molecular dynamics run.
A straightforward way of doing this is to perform an \textit{ab inito} molecular dynamics simulation, and sampling the resulting trajectories. 

All AIMD trajectories were built in the \textit{Vienna Ab Initio Simulation Package} (VASP) \cite{kresse1993ab} within the canonical ensemble, with constant number of atoms $N$, constant volume $V$, and constant temperature $T$ (NVT), the average temperature being fixed using a Nose-Hoover thermostat \cite{nose1984molecular,hoover1985canonical}. The dynamics were performed by solving numerically Newton's equations of motion using the Verlet algorithm in the velocity form with a timestep of $1.5$ fs.
The electronic system was treated with electron-ion interactions represented by the projector augmented wave (PAW) potentials \cite{blochl1994projector,kresse1999ultrasoft} and exchange and correlation effects were taken into account with the Generalized Gradient Approximation (GGA) \cite{wang1991correlation} in the Perdew, Burke, and Ernzerhof (PBE) \cite{perdew1981self} formulation. The cutoff energy in the plane-wave expansion was taken as $300$\;eV. 
For such large supercells and in the liquid state, only the $\Gamma$ point was used for sampling the Brillouin zone. 

The dataset used in this work was constructed in such a way, with configurations drawn from a number of \textit{ab initio} simulations performed specifically for this task.
The majority of these simulations were performed at Ni compositions  $x_{\rm Ni}= 0.25$, $0.5$, and $0.75$, as well as pure Ni, and pure Al, with the aim of covering the entire composition range.
For each composition, two branches of simulations were performed, one starting from a high temperature liquid, quenching in steps of $200$~K, and one branch starting from a $10$~K crystalline state obtained from the \textit{materials project} database \cite{jain2013commentary}, and increasing the temperature in steps of $200$~K.
These simulations were carried out for $60$~ps, with some of the liquid alloy simulations continued up to $90$~ps in order to ensure that the correct chemical ordering appears and is included in the dataset.
A few runs were also performed at a higher pressure, at high temperature, to probe the short-range interaction, and improve the stability of the trained potential. In addition to these primary simulations in the liquid state, additional simulations were performed at $10$~K for the stable Al$_3$Ni$_5$, Al$_4$Ni$_3$, and Al$_2$Ni$_3$ crystalline phases.
The five unstable states on \textit{materials project} closest to the convex hull were included in the same way, namely mp-1183232, mp-1228868, mp-1025044, mp-1229048, and mp-672232. Figure S1 in the supplementary information file shows the $x_\mathrm{Ni}$-T locations of the AIMD trajectories as well as the number of configurations sampled from each of them.

\subsection{High Dimensional Neural Network Potentials}
With the \textit{ab initio} dataset in place, the next step is to define the machine learning model used to construct our MLIP, here the Behler-Parinello HDNNP \cite{behler-parrinello-2007, behler-tutorial-2015}.
In this framework the potential energy of the many-body system is given \textit{via} a nearsightedness approximation,
\begin{equation}
    U\approx\sum_{i=1}^{N}U_i,
\end{equation}
where $N$ is the number of atoms in the system, and $U_i$ gives the local energy contribution of atom $i$, dependent only on atoms $j$ within some neighborhood $r_{ij}=|r_j - r_j| < r_c$ defined by some cutoff distance $r_c$.
In the Behler-Parinello HDNNP approach, $U_i$ is chosen as a species-dependent neural network, taking as input an atomic fingerprint, here given by the Behler-Parrinello Symmetry Functions (SFs)\cite{behler-fingerprints-2011}:
\begin{align}
  \label{eq:G2}
  G^2_i &= \sum_j e^{-\eta(R_{ij}-R_s)^2}f_c(R_{ij})\\
  \label{eq:G5}
  G^5_i &= 2^{1-\zeta}\sum_{j,k}(1 + \Lambda\cos\theta_{ijk})^\zeta
  e^{-\eta(R_{ij}^2+R_{ik}^2+R_{jk}^2)}f_c(R_{ij})f_c(R_{ik})f_c(R_{jk})\;.
\end{align}
Here, $R_{ij}$ is the distance between atoms $i$ and $j$, $\theta_{ijk}$ is the angle between atoms $j$ and $k$ with respect to atom $i$, and $f_c(R_{ij})$ is defined as $0$ for $R_{ij} > r_c$ and for $R_{ij}<r_c$ as a polynomial going smoothly to 0 at the neighborhood cutoff $R_{ij} = r_c$.
The parameters $\eta$, $\zeta$, $\Lambda$, and $R_s$ allow for defining a set of {features} by assigning to  these parameters different values.

Once the number of SFs used for each species, their parameters, as well as the architecture of the NNP of each species, were chosen, the weights and biases of the NNPs are optimized by minimizing the mean-square error of the HDNNP predictions, with respect to the \textit{ab initio} training dataset.
A validation dataset and a test dataset are considered, independent of the training dataset. 
The validation set is used to evaluate the extrapolation error, to guide the tuning of hyperparameters, and to decide when to end training as part of early stopping.
The test dataset is then used to evaluate the final model, after training is completed. All the settings of the HDNN and the parameters of the SFs can be found in \cite{Sandberg2024Materialscloud}.

While the HDNNP fits the potential energy surface, it is often beneficial to fit not only the potential energy, but also its gradients, namely the \textit{ab initio} forces.
In this work, however, only the energy was considered for  the fitting of our HDNNP model.
This is to maintain continuity with our previous work on the solidification of pure Al, as well as to not potentially reduce the accuracy of the energetics, which play an important role in the phenomenon of nucleation.
This will also allow for a more straightforward future extension of our previously proposed embedded feature selection method \cite{sandberg2024feature} to this multicomponent system.
Training itself was performed using the N2P2 package \cite{singraber2019parallel,singraber2019library}.

\subsection{Classical simulations and analysis}

Classical simulations were performed in LAMMPS \cite{thompson2022lammps}, using the \textit{ml-hdnnp} plugin provided by N2P2 \cite{singraber2019library}.
Time integration was performed using the velocity Verlet algorithm, with a timestep of $1$ fs.
Simulations for validating the potential were performed in the NVT ensemble, with temperature fixed by a Nos\'{e}-Hoover thermostat, and volume pressure controlled indirectly by fixing the volume.
MD simulations of homogeneous nucleation were performed in the NPT ensemble, adding a barostat fixing the pressure to ambient conditions.
Visualization of the system, common neighbor analysis, and calculation of the pair-correlation functions were performed using OVITO \cite{stukowski2009visualization}.
Calculation of structure factors were performed using ISAACS \cite{le2010isaacs}, and the bond-order parameters were calculated with pyscal \cite{menon2019pyscal}.

\subsection{Structural and dynamic properties}

Basic structural information of the melt can be obtained from the pair-correlation functions (PCF) \cite{young1992structural,hansen2006theory}. The partial PCF, $g_{ij}(r)$, gives the probability of finding a particle of type $i$ at a distance $r$ from a particle of type $j$, and can be obtained from simulation by measuring the number of $j$ particles in a spherical shell of thickness $\Delta r$ and radius $r$.
The ensemble average $n_{ij}(r)$ then gives
\begin{equation}
    g_{ij}(r) = \frac{N_{j}}{V}\frac{n_{ij}(r)}{4\pi r^2 \Delta r},
    \label{eq:rdf}
\end{equation}
with $V$ the volume of the simulation box, containing $N_{j}$ total number of $j$ particles. A Fourier Transform of these partials leads to the partial structure factors
\begin{equation}
    S_{ij}(q) = \frac{1}{N}\left<\sum_{k=1}^{N_i}\sum_{l=1}^{N_j}
    \exp \left[i\mathbf{q}\cdot (\mathbf{r}_k^i - \mathbf{r}_l^j)\right]\right>,
    \label{eq:S}
\end{equation}
written here in the Faber-Ziman form \cite{young1992structural}. Weighting the $S_{ij}(q)$ with appropriate scattering lengths from neutron or x-ray diffraction leads to the corresponding total Structure factors $S_N(q)$ or $S_X(q)$. 

The self-diffusion of atoms in the melt is obtained \textit{via} the Mean Squared Displacement (MSD) at time $t$ given by
\begin{equation}
    R^2_i(t) = \frac{1}{N_i}\sum_j^{N_i} \left<\left[\mathbf{r}_j(t+t_0) - \mathbf{r}_j(t_0)\right]^2\right>,
    \label{eq:msd}
\end{equation}
where the the sum runs over atoms of species $i$, $\mathbf{r}_j(t)$ denotes the position of atom $j$ at time $t$, and $<.>$ averages over the initial timestep $t_0$.
From this quantity, the self-diffusion $D_i$ can be extracted via the Einstein relation
\begin{equation}
    D_i = \lim_{t\rightarrow\infty} \frac{R^2(t)}{6t}.
    \label{eq:selfdiffusion}
\end{equation}
The shear viscosity of the system is calculated via a Green-Kubo relation from the stress-stress autocorrelation function
\begin{equation}
    \eta = \frac{V}{k_BT}\int_0^\infty dt \left<P_{xy}(t)P_{xy}(0)\right>,
    \label{eq:viscosity}
\end{equation}
with $V$ the volume of the system, $T$ the temperature, $k_B$ the Boltzmann constant, and $P_{xy}(t)$ the off-diagonal component of the stress tensor at time $t$.
Note that, for an isotropic system, we can average over the $xy$, $xz$, and $yz$ components in evaluating the ensemble average.

\section{Results and discussion}
\label{sec:results}
\subsection{Structure and Dynamics}

\begin{figure}
    \centering
    \includegraphics[width=0.66\textwidth]{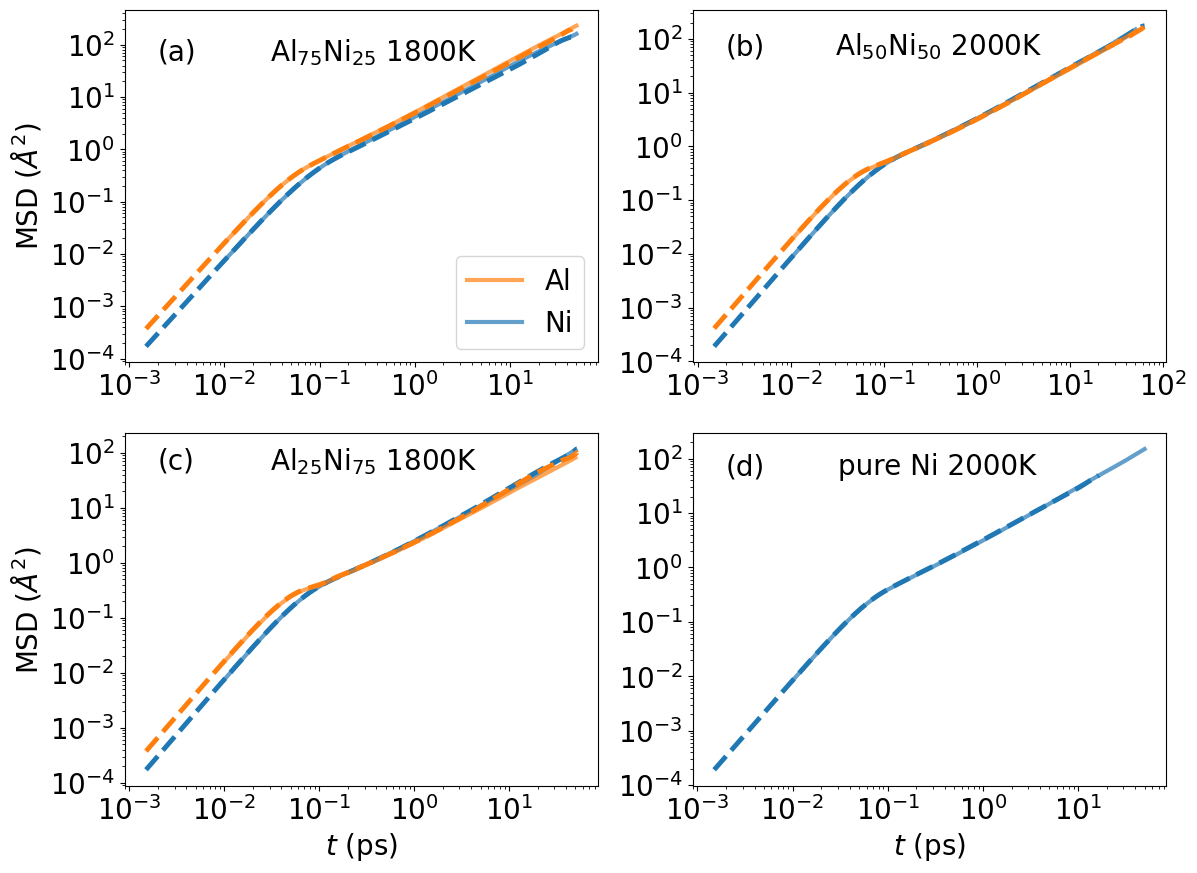}
    \includegraphics[width=0.33\textwidth]{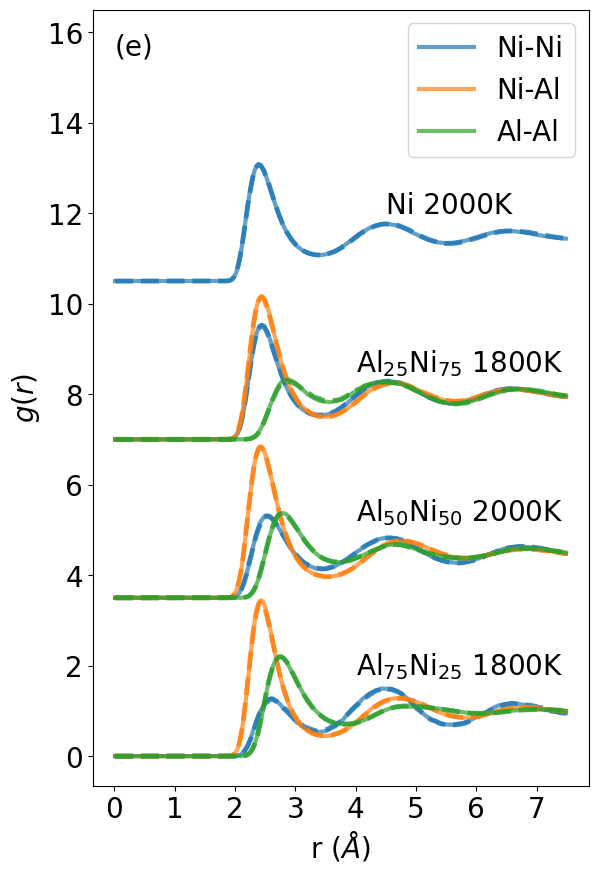}
    \caption{Comparison between NNP and AIMD results for the melt.
    NNP results are shown as solid lines, AIMD as dashed lines.
    MSD values are for (a) $1800$ K for Al\textsubscript{75}Ni\textsubscript{25}, (b) $2000$ K for Al\textsubscript{50}Ni\textsubscript{50}, (c) $1800$ K for Al\textsubscript{25}Ni\textsubscript{75}, and (d) $2000$ K for pure Ni.
    PCFs are shown in (e) for the same temperatures, with subsequent compositions shifted upwards by $3.5$, $7$ and $10.5$.}
    \label{fig:aimd-comparison}
\end{figure}

\begin{table}
    \centering
    \caption{Self diffusion coefficients extracted from the mean square displacements in Figure \ref{fig:aimd-comparison} for each Ni composition, $x_{Ni}$. Values in parentheses are from AIMD.}
    \begin{tabular}{ccccc}
        \hline\hline
         $x_{Ni}$ & $0.25$ & $0.50$ & $0.75$ & $1.0$  \\
         $T$ (K) & $1800$ & $2000$ & $1800$ & $2000$  \\
         $D_\mathrm{Ni}\;[10^{-9}m^2/s]$ & $5.108\;(5.578)$ & $5.003\;(4.858)$ & $3.884\;(3.640)$ & $4.980\;(4.977)$\\
         $D_\mathrm{Al}\;[10^{-9}m^2/s]$ & $7.598\;(7.414)$ & $4.527\;(4.316)$ & $2.646\;(3.043)$ & -\\
        \hline\hline
        
    \end{tabular}
    \label{tab:diffusion}
\end{table}

This section is devoted to the validation of the trained HDNNP on the structural properties, such as the  pair-correlation functions and the total and partial structure factors, as well as the dynamics through the self-diffusion coefficients and the viscosity. 
Comparison with AIMD assesses the quality of the training procedure, while comparison with experiments tests the reliability of the \textit{ab initio} simulation scheme. 

As a first evaluation of the potential, the HDNNP is compared to \textit{ab initio} simulations, by performing a set of simulations with $N=256$ atoms corresponding to the AIMD system size (to exclude finite-size effects in the comparison).
A set of configurations were taken from the end of the \textit{ab initio} trajectories as an initial state.
The temperature, matching that used in \textit{ab intio}, was chosen to be in the liquid, just above the experimental liquidus line.
After initializing the atom velocities according to a Maxwell-Boltzmann distribution, an equilabration of $100$ ps was performed, followed by a $200$ ps production run.

The mean-square displacements are shown in Figure \ref{fig:aimd-comparison} (a)-(d), along with those calculated from the \textit{ab initio} trajectories.
The corresponding values for the self-diffusion coefficients, obtained from the MSD \textit{via} the Einstein relation given by Eq. (\ref{eq:selfdiffusion}) for both the NNP and for \textit{ab initio} are in good agreement, as shown in Table \ref{tab:diffusion}.
Note that the largest disagreement occurs for the minority components, Al in Al\textsubscript{25}Ni\textsubscript{75} and Ni in Al\textsubscript{75}Ni\textsubscript{25}, with other components typically agreeing to within roughly $0.2\times10^{-9}m^2/s$, which can be seen as a reasonably good agreement. 

The partial pair-correlation functions are displayed in Figure \ref{fig:aimd-comparison} (e), and it can be seen that the NNP curves show a nearly perfect match with the AIMD ones. 
Especially, it reproduces the quite strong affinity between Al and Ni revealed through the large amplitude of the first peak of the Al-Ni partial with respect to the Al-Al and Ni-Ni ones. 
This indicates a chemical short-range ordering \cite{jakse2015relationship} that has a maximum for Al\textsubscript{50}Ni\textsubscript{50}. 
Table SI of the supplementary information file (SI) shows that for this composition the first peak maximum of the Al-Ni partial is the maximum. All these results highlight the high quality of the training procedure in the whole range of compositions.

\begin{figure}
    \centering
    \includegraphics[width=0.48\textwidth]{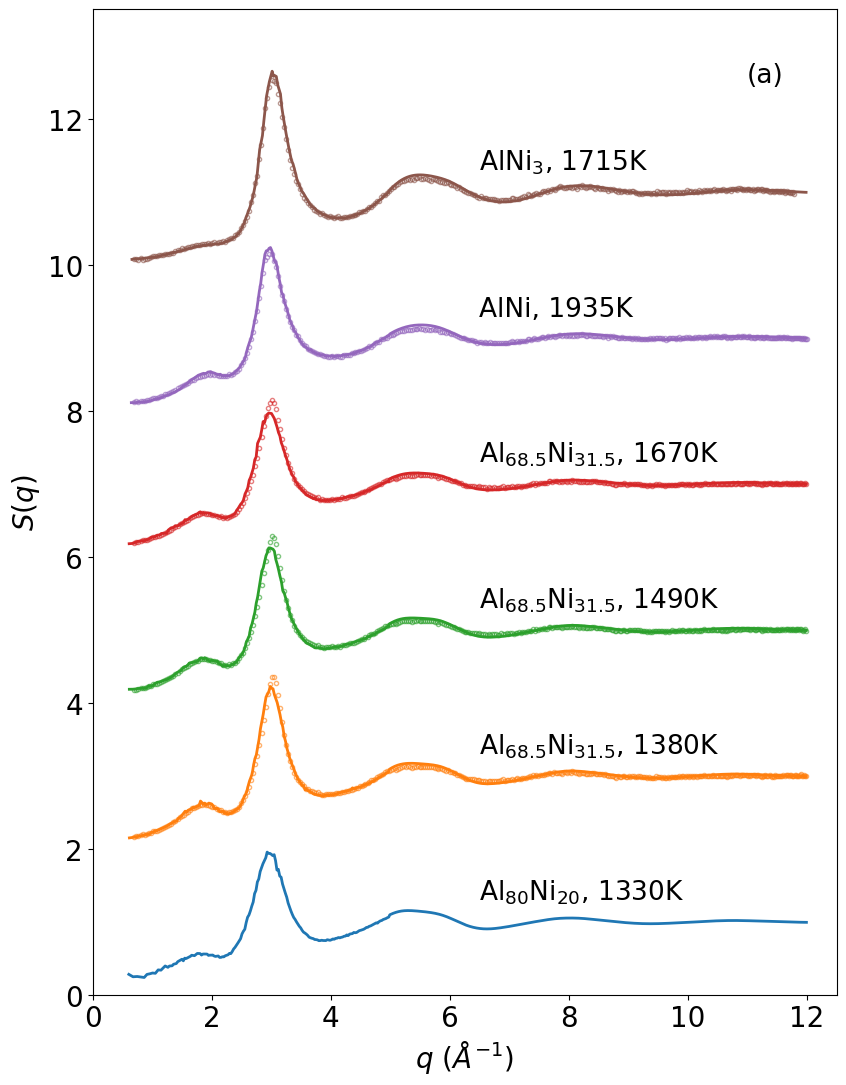}
    \includegraphics[width=0.48\textwidth]{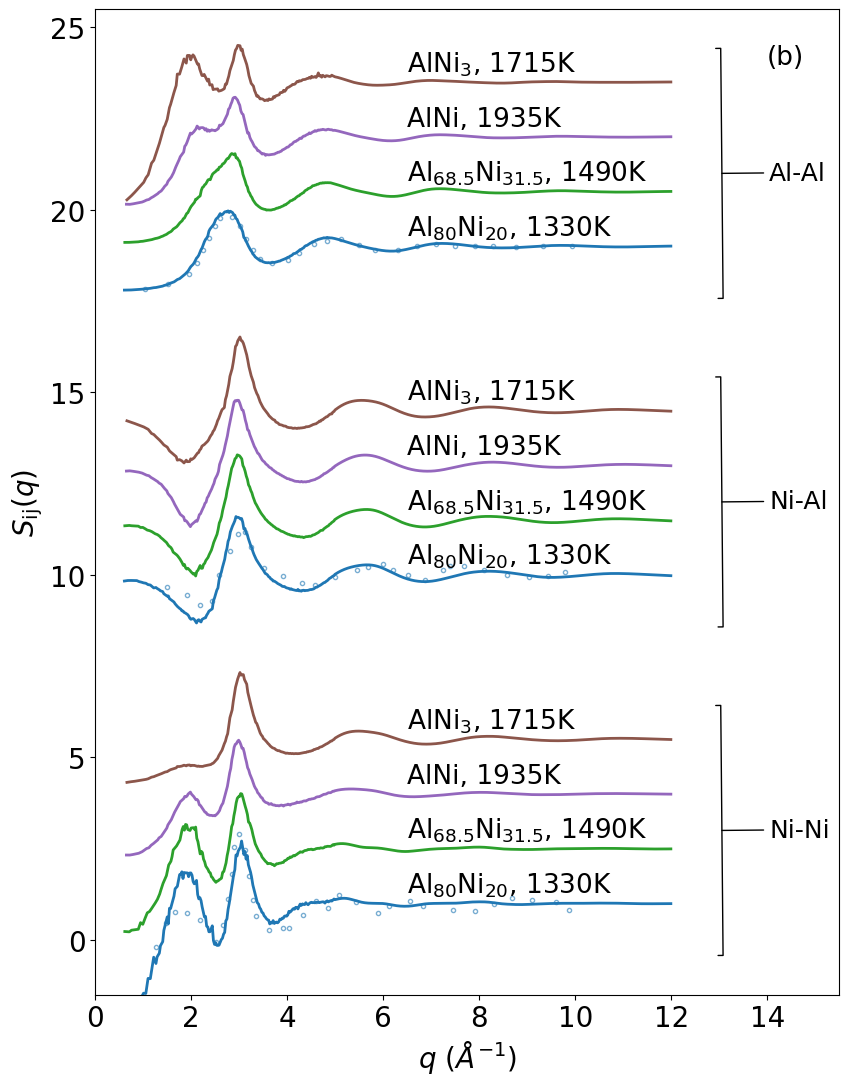}
    \caption{Total structure factor from neutron diffraction for liquid Al-Ni above the liquidus (a), and corresponding Faber-Ziman partial structure factors (b). Comparison between HDNNP (solid lines) and experimental total structure factor \cite{belova2019computer} (symbols) for Al\textsubscript{25}Ni\textsubscript{75} at $1715$\;K,  Al\textsubscript{50}Ni\textsubscript{50} at $1935$\;K, Al\textsubscript{68.5}Ni\textsubscript{31.5} at $1700$, $1490$, and $1380$\;K. and Al\textsubscript{80}Ni\textsubscript{20} at $1935$\;K. The curves are shifted upwards by an amount of 2. The partial structure factors Al\textsubscript{80}Ni\textsubscript{20} was compared to the experimental data of \cite{maret1990structure}. The different partials are shifted by an amount of 9 and subsequently shifted by 1.5 for the compositions.}
    \label{fig:Stot}
\end{figure}

A comparison with experimental data is now done through the total and partial structure factors, the temperature and composition dependence self-diffusion coefficients as well as the viscosity. Figure \ref{fig:Stot}(a) shows the structure factor from the HDNNP and compared to the recent neutron diffraction experiments of Belova \textit{et al.} \cite{belova2019computer} and Maret \textit{et al.} \cite{maret1990structure}. 
An excellent agreement is found for the peak positions as well as their amplitudes for all the considered compositions. 
The partials structure factors in the Faber-Ziman formulation \cite{young1992structural} are shown in Figure \ref{fig:Stot}(b) for the four different compositions. 
For Al\textsubscript{80}Ni\textsubscript{20}, the partials agree remarkably well with the data of Maret et al. \cite{maret1990structure}. 
Another interesting feature is that the amplitude of the pre-peak of the Ni-Ni partial decreases with increasing Ni composition and is seen to correlate tightly with that of the total structure factor. 
This suggests that the pre-peak of the total structure factor might originate from the Ni-Ni structural features. 
It was argued this pre-peak could be a signature of the chemical short-range order \cite{brillo2006local}, and it was later found that it might be characteristic of a medium range ordering \cite{roik2010medium,jakse2017interplay}.  


\begin{figure}
    \centering
    \includegraphics[width=0.48\textwidth]{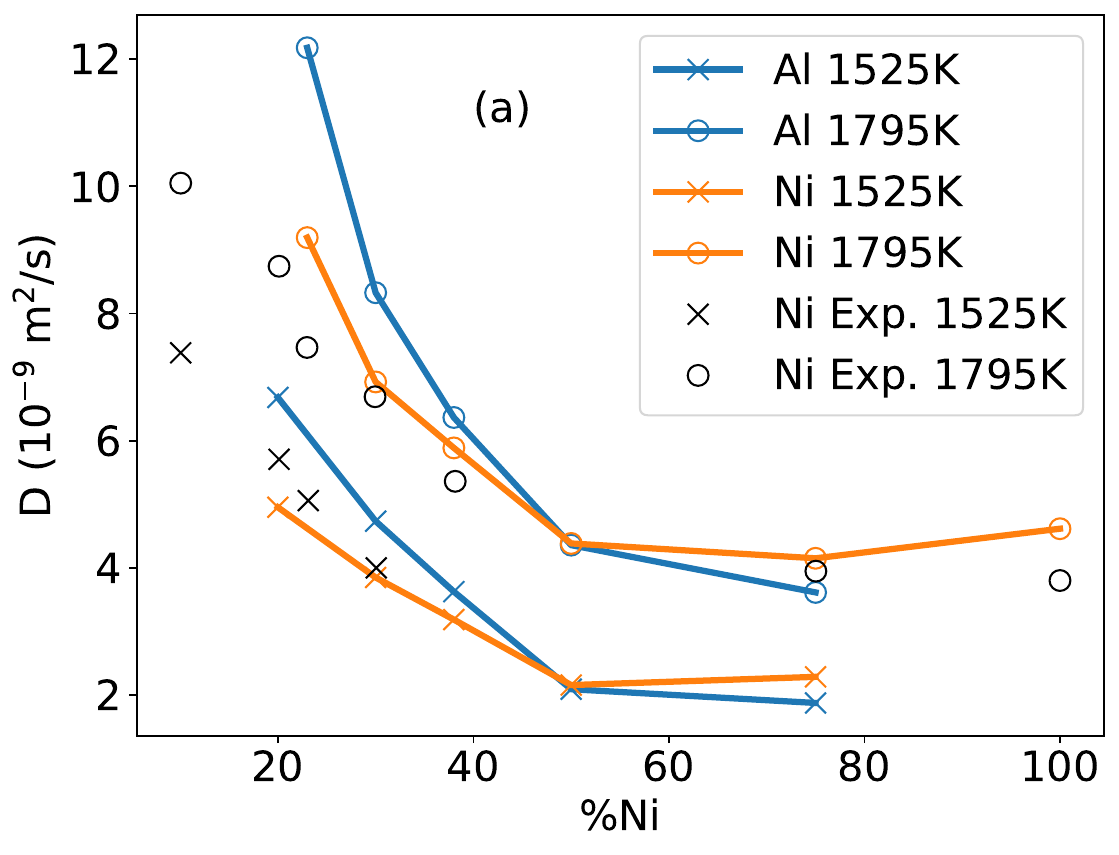}
    \includegraphics[width=0.48\textwidth]{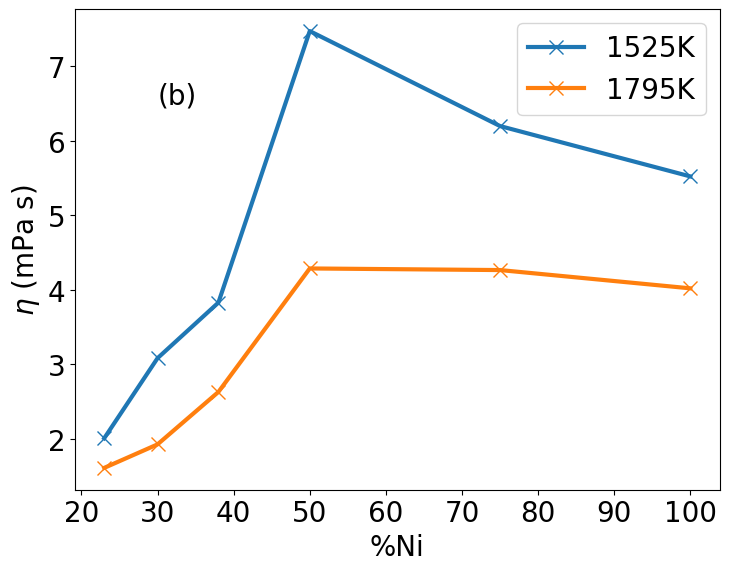}
    \caption{Composition dependence of self-diffusions (a), and viscosity (b), along $1795$K and $1525$K isotherms.
    Solid lines show results of this article, with dashed lines showing \textit{ab initio} data of \cite{jakse2015relationship}, and separate symbols showing experimental data from \cite{das2005influence}.}
    \label{fig:self-iso}
\end{figure}

\begin{figure}
    \centering
    \includegraphics[width=0.48\textwidth]{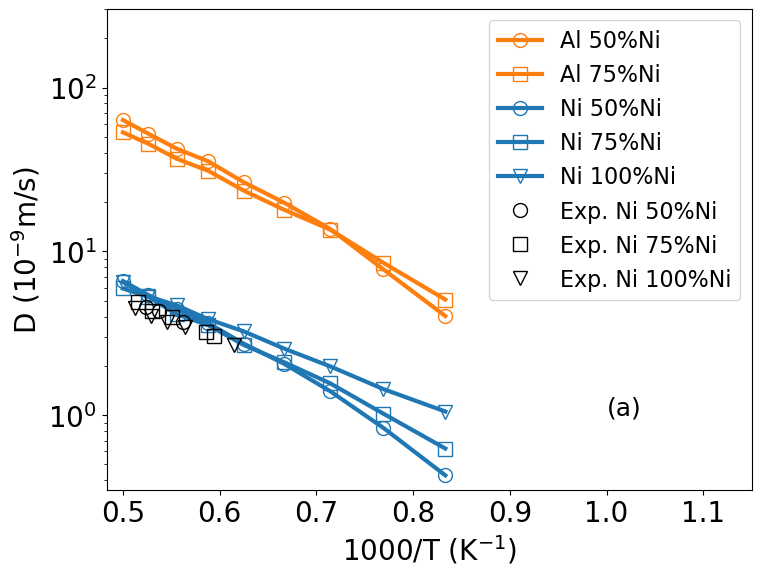}
    \includegraphics[width=0.48\textwidth]{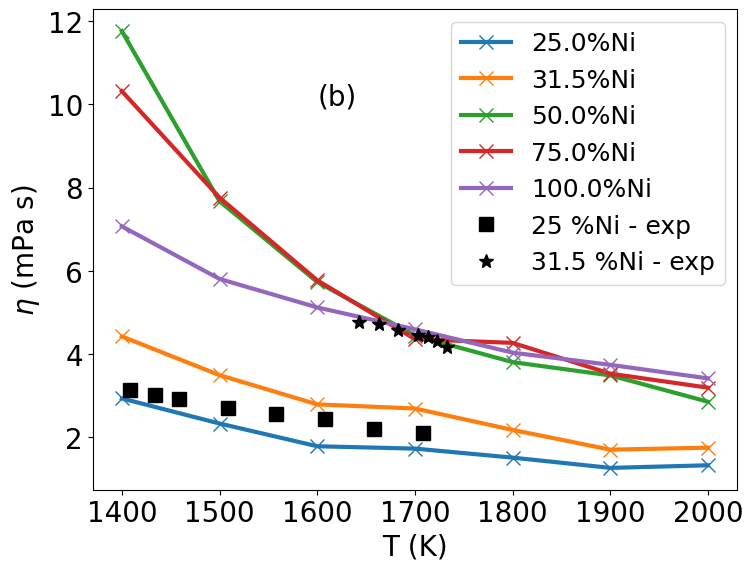}
    \caption{Composition evolution of self-diffusion coefficients (a) and viscosity (b) at ambient pressure for two selected temperatures.
    Separate symbols show experimental diffusion data from \cite{stuber2010ni}, and experimental viscosities from \cite{egry2010thermophysical}.}
    \label{fig:self-arrhenius}
\end{figure}

The self-diffusion behaviour with respect to the composition is examined, seeking to match the experimental data points in \cite{das2005influence}.
At each composition the system is initiated in a $13500$ atom fcc crystal, with atoms randomly assigned to each species, in the given ratio.
The system is then melted at a high temperature of roughly $2000$~K for $100$~ps, then quenched to the final temperature over $10$~ps, and equilibrated for $100$~ps, all in a NPT ensemble.
Measurements are then performed over $120$~ps in a NVT ensemble.
Figure \ref{fig:self-iso}(a) shows the resulting self-diffusion coefficients, calculated via the Einstein relations. 
The agreement with the experimental values, shown as separate symbols in the plot, is quite good.
A progressive decoupling of the two self-diffusion coefficients sets in as the Ni content decreases above $x_{Ni}=0.50$. A similar trend was observed in previous AIMD simulations \cite{jakse2015relationship}.
The same simulation procedure is repeated, in a smaller system of $500$ atoms to calculate the viscosity.
The latter was calculated \textit{via} the Green-Kubo relation given in Eq. (\ref{eq:viscosity}), for which each simulation is repeated $10$ times to obtain a better ensemble averaging in the calculation of the stress-stress autocorrelation function.
Figure \ref{fig:self-iso}(b) displays its composition dependence for the two selected temperatures, which shows consistently an opposite trend to the self-diffusion, as expected. 
It worth mentioning that by decreasing the temperature, a clear maximum occurs at equiatomic composition. 
This can be understood by the larger Al-Ni affinity for this composition (see Table SI in the SI file) that promotes the compound formation. 

A second set of simulations was performed to study the temperature dependence.
After being initialized as for the previous set of simulations, the system was quenched in steps of $100$~K.
At each temperature point a $20$~ps equilibration is performed, followed by a $60$~ps measurement.
The temperature dependence of the self-diffusion coefficients is shown in Figure \ref{fig:self-arrhenius}(a).
Again, a smaller system is simulated with the same procedure, and used to calculate the viscosities, averaging the stress-stress autocorrelation over $10$ realizations before applying the Green-Kubo relations.
The viscosities are shown in Figure \ref{fig:self-arrhenius}(b).
The simulations for the viscosities are also performed at two additional compositions, to compare with existing experimental values of \cite{egry2010thermophysical}, showing a good agreement for Al\textsubscript{75}Ni\textsubscript{25}, but an underestimation at Al\textsubscript{68.5}Ni\textsubscript{31.5}.

\subsection{ Homogeneous Nucleation}
\begin{figure}
    \centering 
    \vcenteredhbox{\begin{minipage}{0.19\textwidth}
		{\large \hfill{$t = 170$~ps}\hfill}
    \end{minipage}}	
    \vcenteredhbox{\begin{minipage}{0.19\textwidth}
		{\large \hfill{$t = 220$~ps}\hfill}
    \end{minipage}}
    \vcenteredhbox{\begin{minipage}{0.19\textwidth}
		{\large \hfill{$t = 270$~ps}\hfill}
    \end{minipage}}
    \vcenteredhbox{\begin{minipage}{0.19\textwidth}
		{\large \hfill{$t = 320$~ps}\hfill}
    \end{minipage}}
    \vcenteredhbox{\begin{minipage}{0.19\textwidth}
		{\large \hfill{$t = 500$~ps}\hfill}
    \end{minipage}}
    
    \includegraphics[width=0.19\textwidth]{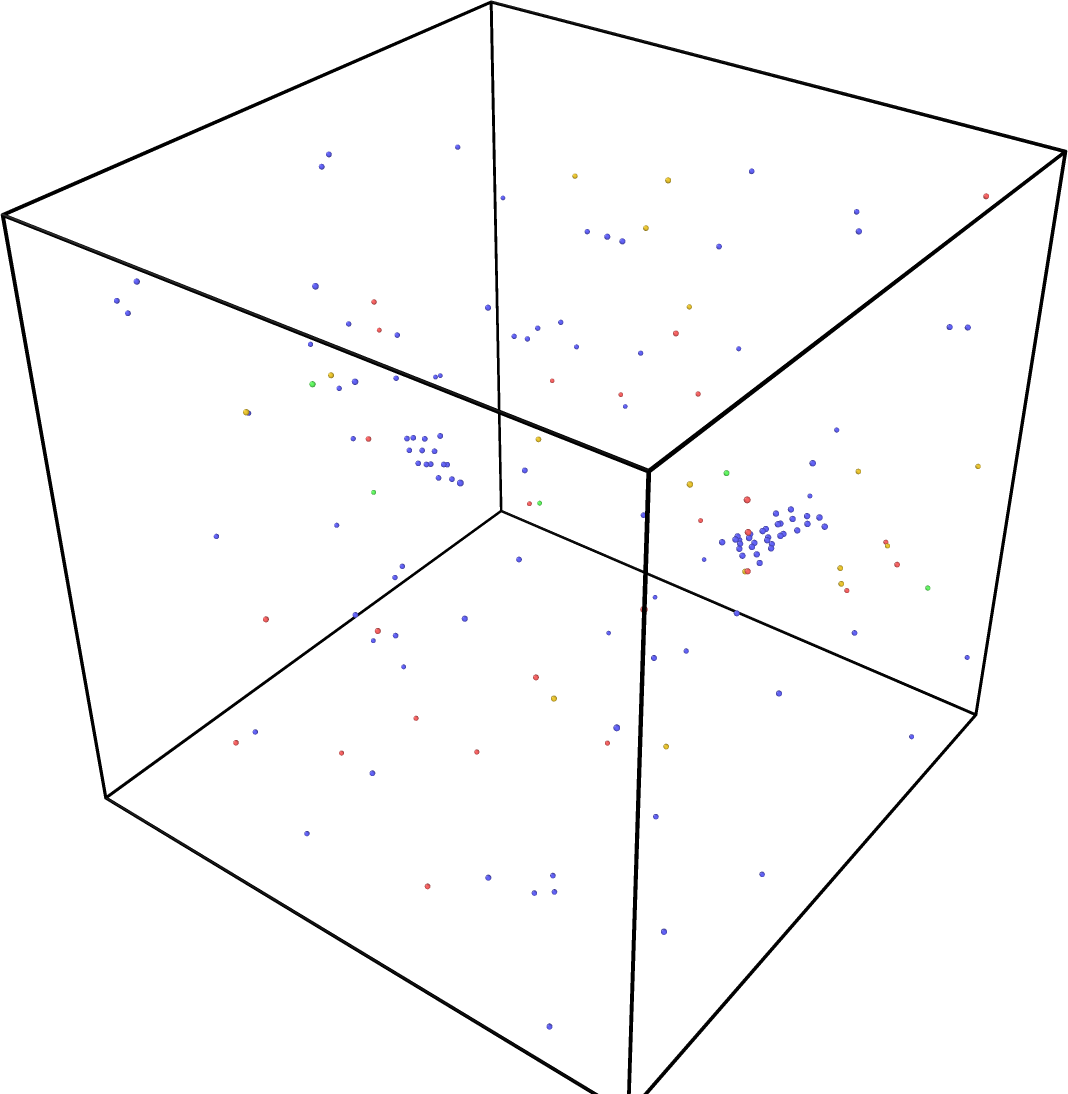}
    \includegraphics[width=0.19\textwidth]{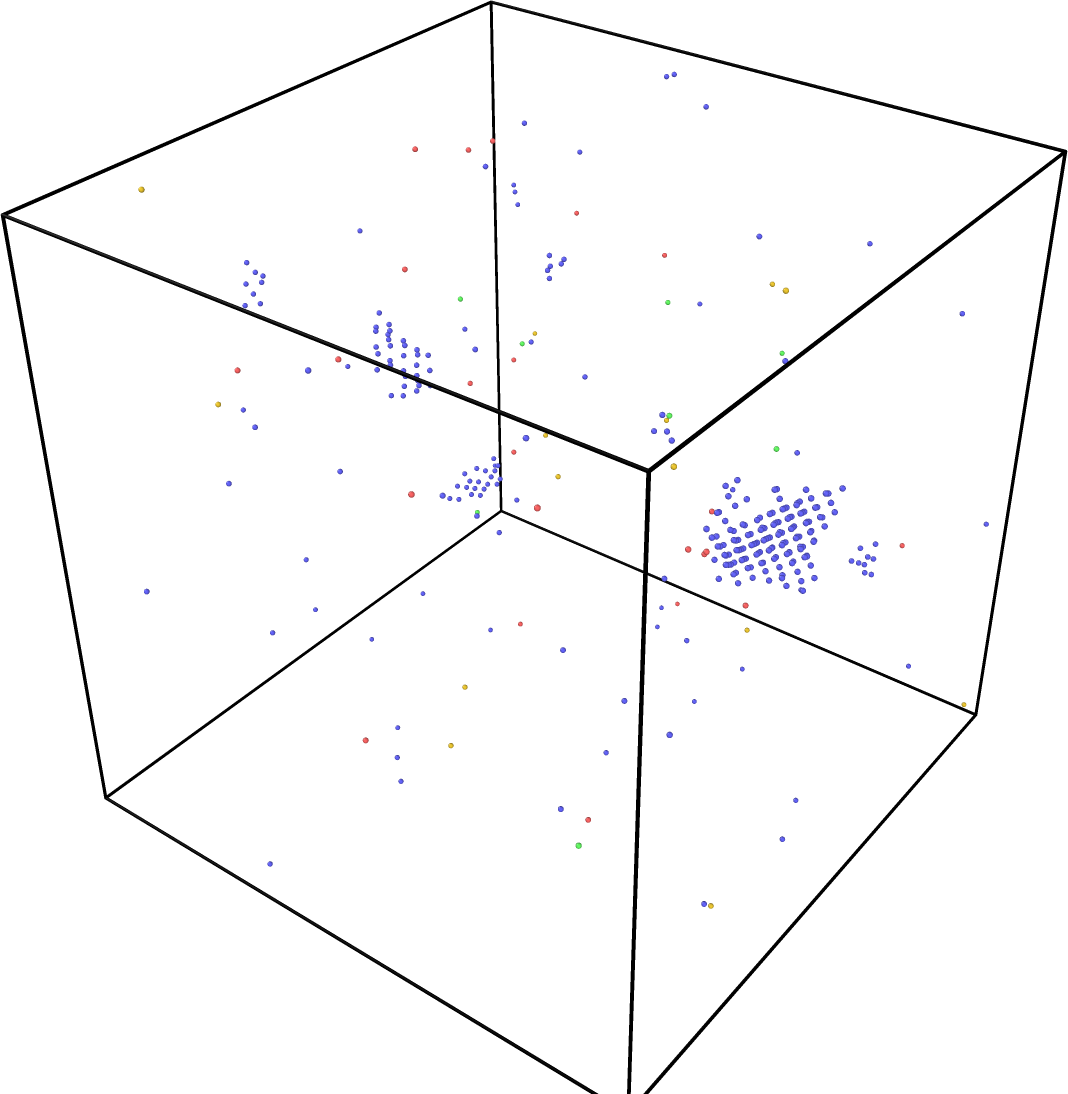}
    \includegraphics[width=0.19\textwidth]{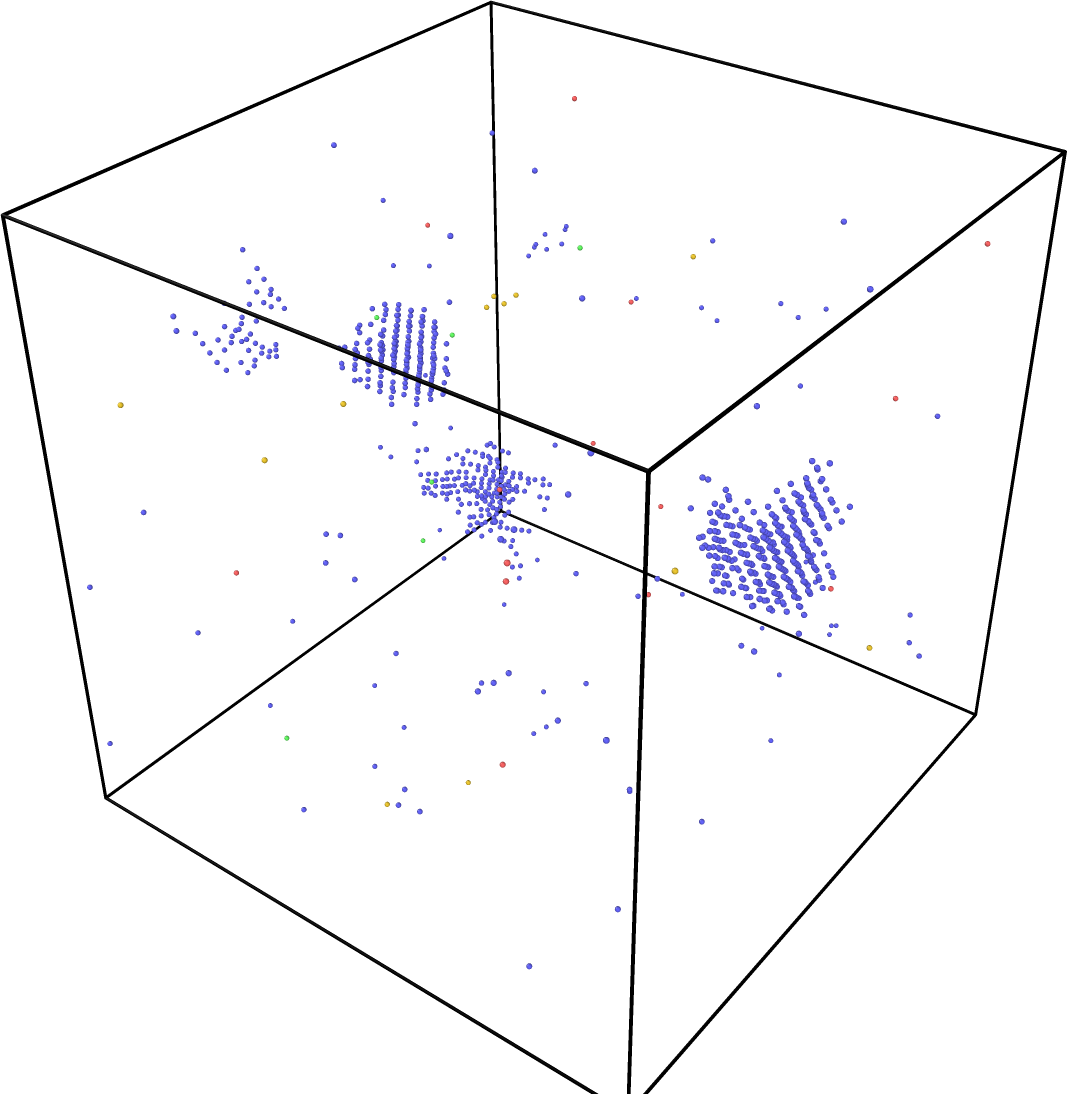}
    \includegraphics[width=0.19\textwidth]{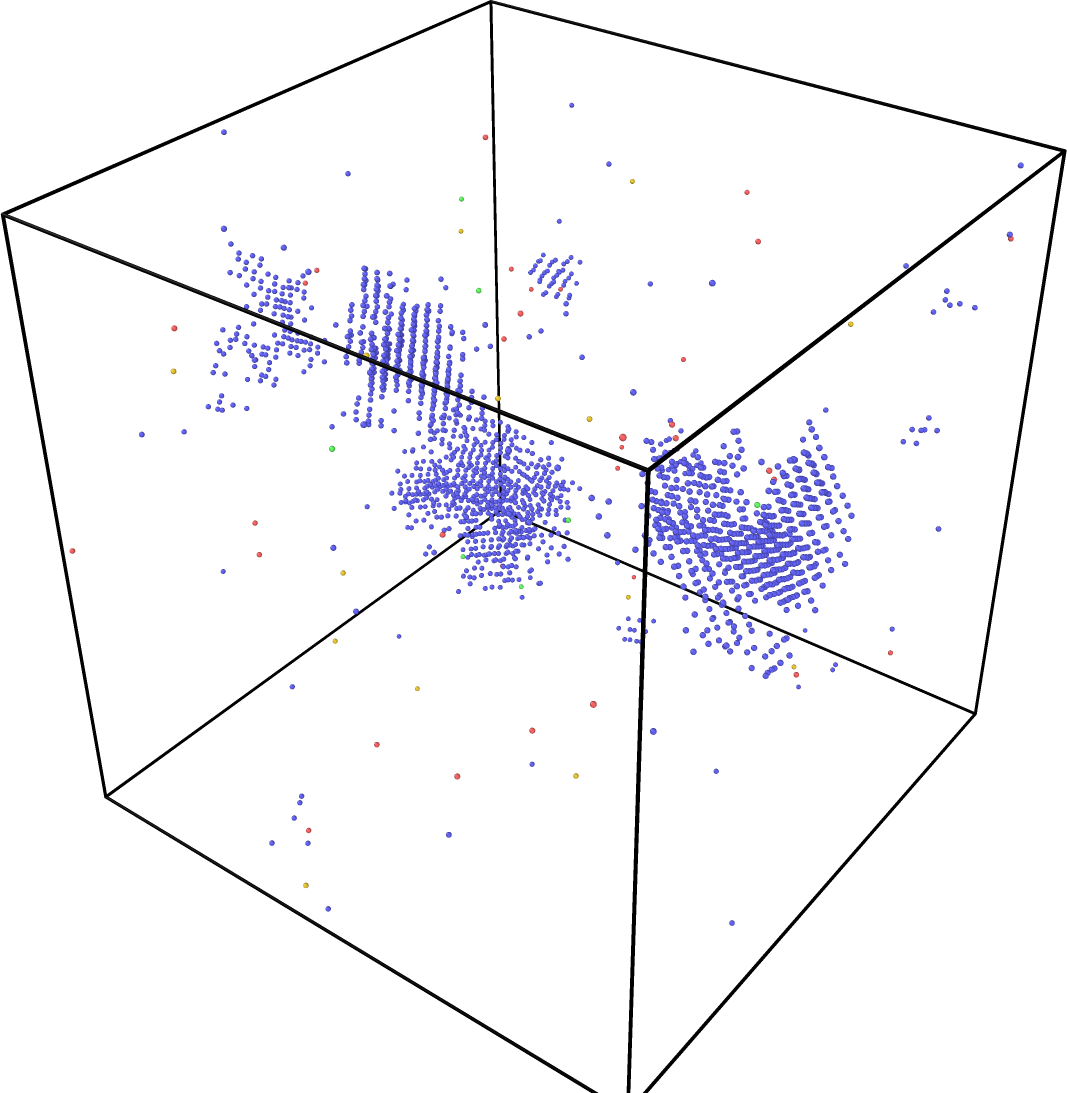}
    \includegraphics[width=0.19\textwidth]{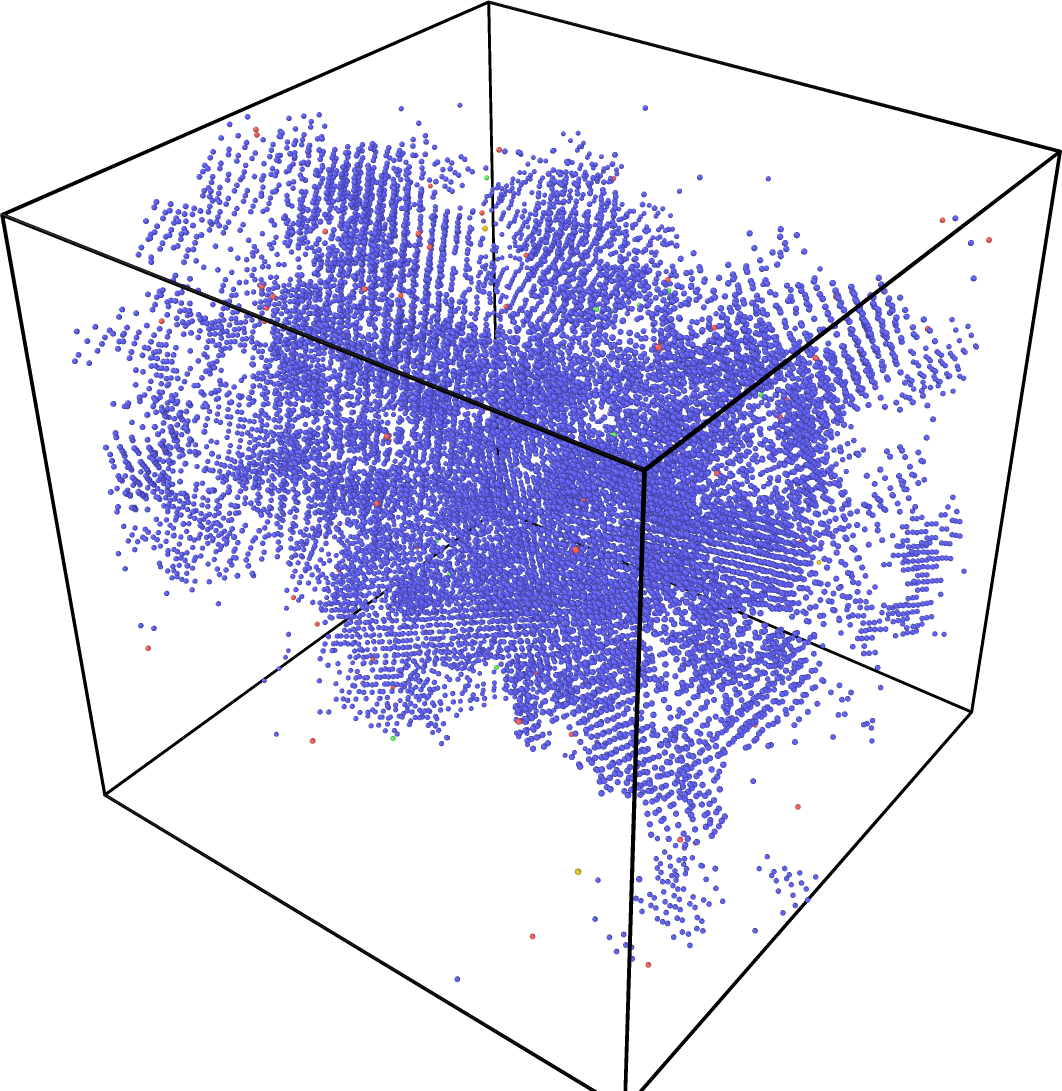}
    \caption{Snapshots of the AlNi system during nucleation, at $170$, $220$, $270$, $320$, and $500$ ps. Particles are colored according to their CNA signature, blue for bcc, red for hcp, green for fcc, and unidentified atoms not shown.}
    \label{fig:snapshots}
\end{figure}

With our ML potential at hand, the early stages of crystal nucleation of Al-Ni alloy at equiatomic composition as well as pure Ni are investigated  at an unprecedented accuracy. 
For this purpose, large scale simulations of $135000$ atoms were performed.
The system is initiated from an fcc solid solution assigning randomly the type of atoms to reproduce Al\textsubscript{50}Ni\textsubscript{50} composition. 
The system is then melted at $2000$K, and equilibrated for $100$ ps, before being quenched to $1300$K with a cooling rate of $13$ K/ps, for an undercooling of $\Delta T \approx 650$ K, compared to the experimental liquidus temperature \cite{okamoto1993ni}.
The simulation is then continued at this deep undercooling, up to $750$ ps.
All of these simulations were performed at ambient pressure, within an NPT ensemble to observe the onset of nucleation as can be seen in Figure~\ref{fig:snapshots}, showing a sequence of snapshots of the simulation box at different times. 

\begin{figure}
    \centering 
    \includegraphics[width=0.4\textwidth]{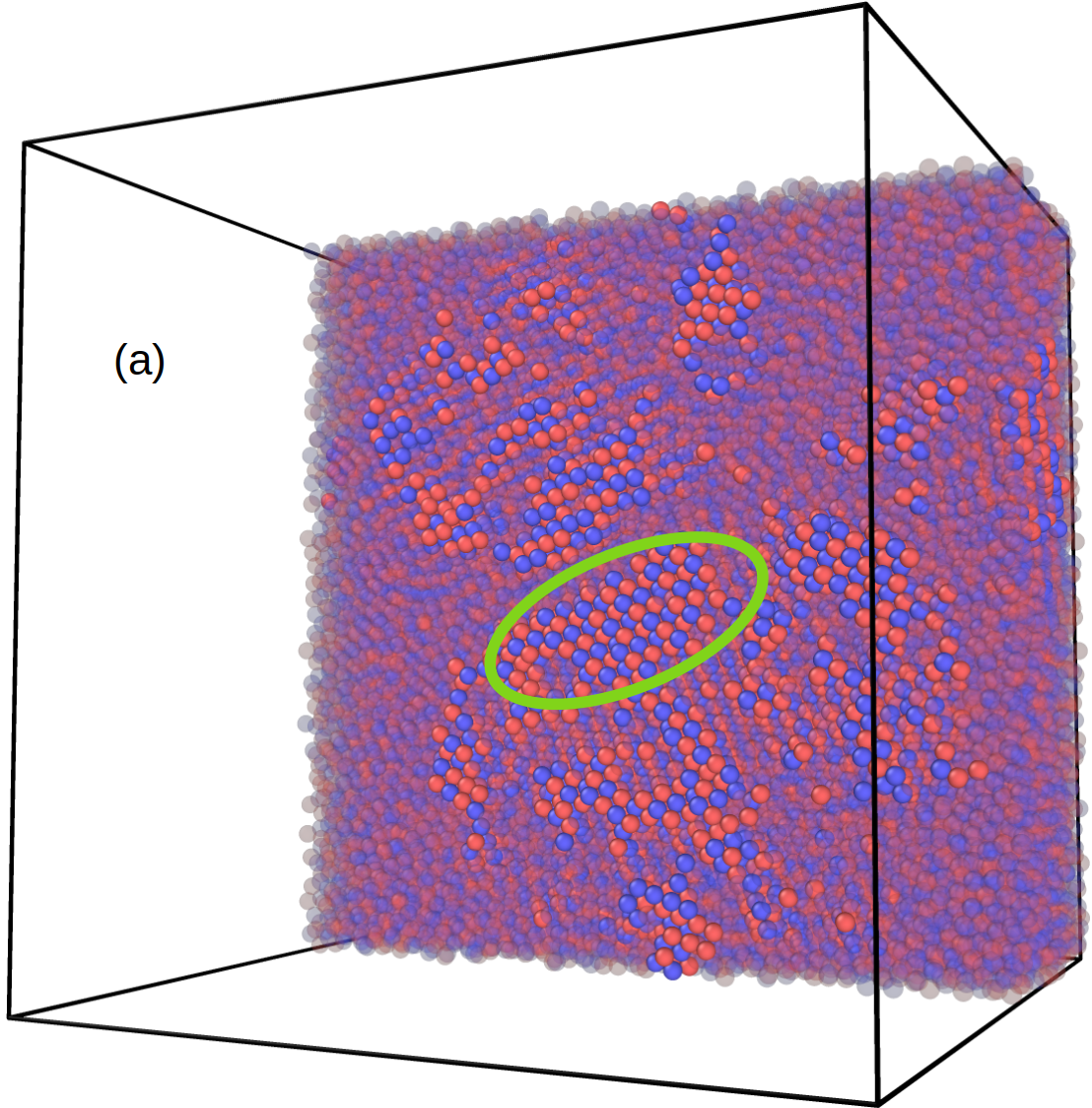}
    \includegraphics[width=0.5\textwidth]{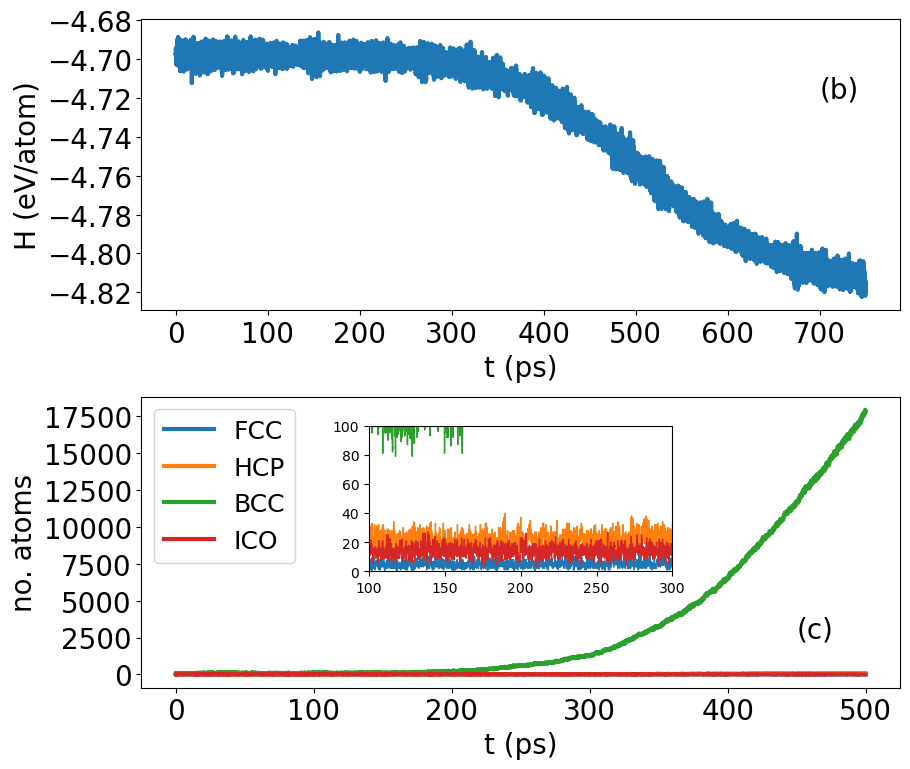}
    \caption{Slice of the simulation box for AlNi during nucleation at 500 ps showing the inner structure of the nuclei in the B2 Structure (a). Enthalpy as a function of time during nucleation (b). Abundance of local structures, identified with the CNA, in the AlNi system (c).}
    \label{fig:timeseries:50}
\end{figure}

For all the snapshots, a common neighborhood analysis is performed to classify the local topology of atoms.
To reduce the influence of thermal noise, before analyzing the configurations, a minimization is applied on each configuration so as to obtain the inherent structures.
After an incubation period during which many pre-critical nuclei form and dissolve back into the liquid state, the majority are identified as bcc clusters (colored blue), with only sporadic clusters identified as fcc (green) and hcp (red). Atoms identified as neither of these structures are not shown.
Based on visual inspection, we observe nuclei of sizes up to $80$ atoms which disappear back into the liquid, putting a lower bound on the critical size $n^\ast$.
Conversely, the supercritical nuclei appears to arise from clusters to around $120$ atoms representing an upper bound to $n^\ast$.

Nucleation starts around $170$ ps with the growth of a first supercritical nuclei followed by a second one at $220$ ps, which is also detected by a sharp drop in the time evolution of enthalpy drawn in Figure~\ref{fig:timeseries:50}(b). 
The nuclei display directly a bcc-type from the beginning of the nucleation as quantified in Figure~\ref{fig:timeseries:50}(c), while the fcc and hcp ordering remains always marginal. 
This indicates a single-step nucleation process directly from the undercooled liquid into the bcc phase. A closer analysis of the inner part of the nuclei as shown in Figure~\ref{fig:timeseries:50}(a) indicates clearly a B2 structure.
The supercritical nuclei in Figure \ref{fig:snapshots} also shows an apparent non-spherical shape, in contrast to what is assumed in classical nucleation theory \cite{sosso2016crystal}.

\begin{figure}
    \centering
    \vcenteredhbox{\begin{minipage}{0.24\textwidth}
		{\large \hfill{$t = 1.764$~ns}\hfill}
    \end{minipage}}	
    \vcenteredhbox{\begin{minipage}{0.24\textwidth}
		{\large \hfill{$t = 1.77$~ns}\hfill}
    \end{minipage}}
    \vcenteredhbox{\begin{minipage}{0.24\textwidth}
		{\large \hfill{$t = 1.79$~ns}\hfill}
    \end{minipage}}
    \vcenteredhbox{\begin{minipage}{0.24\textwidth}
		{\large \hfill{$t = 1.83$~ns}\hfill}
    \end{minipage}}
    
    \includegraphics[width=0.24\textwidth]{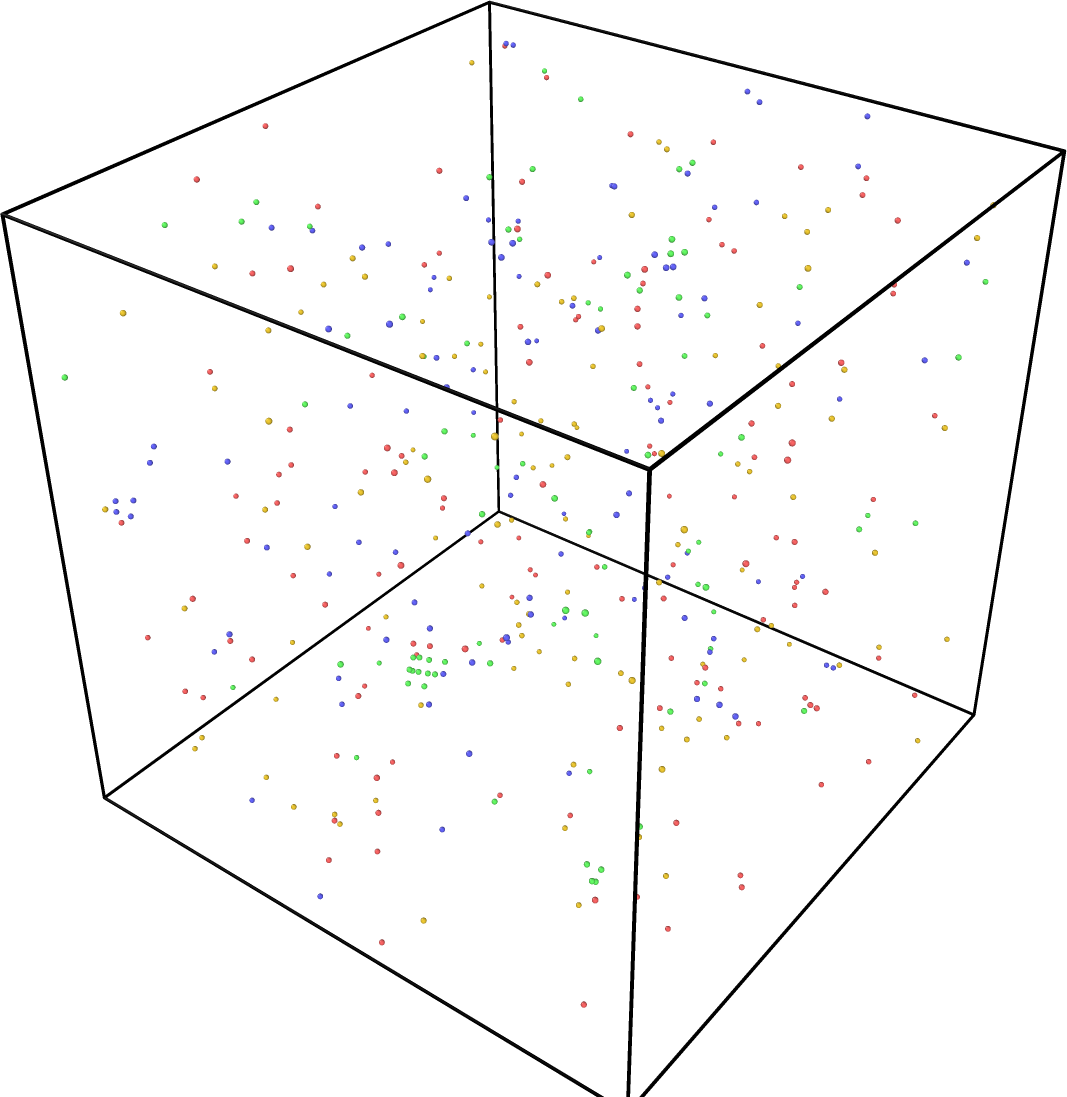}
    \includegraphics[width=0.24\textwidth]{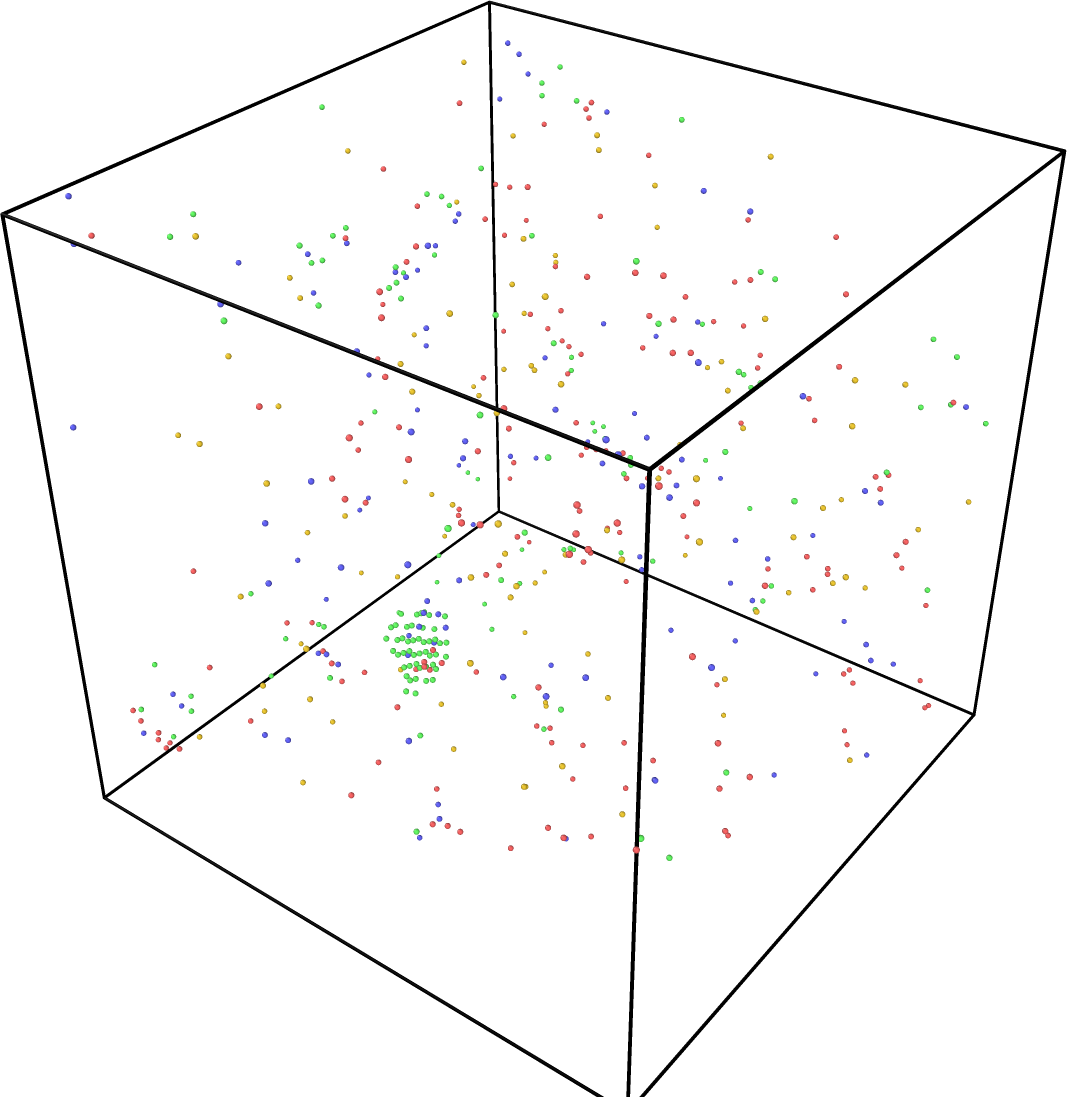}
    \includegraphics[width=0.24\textwidth]{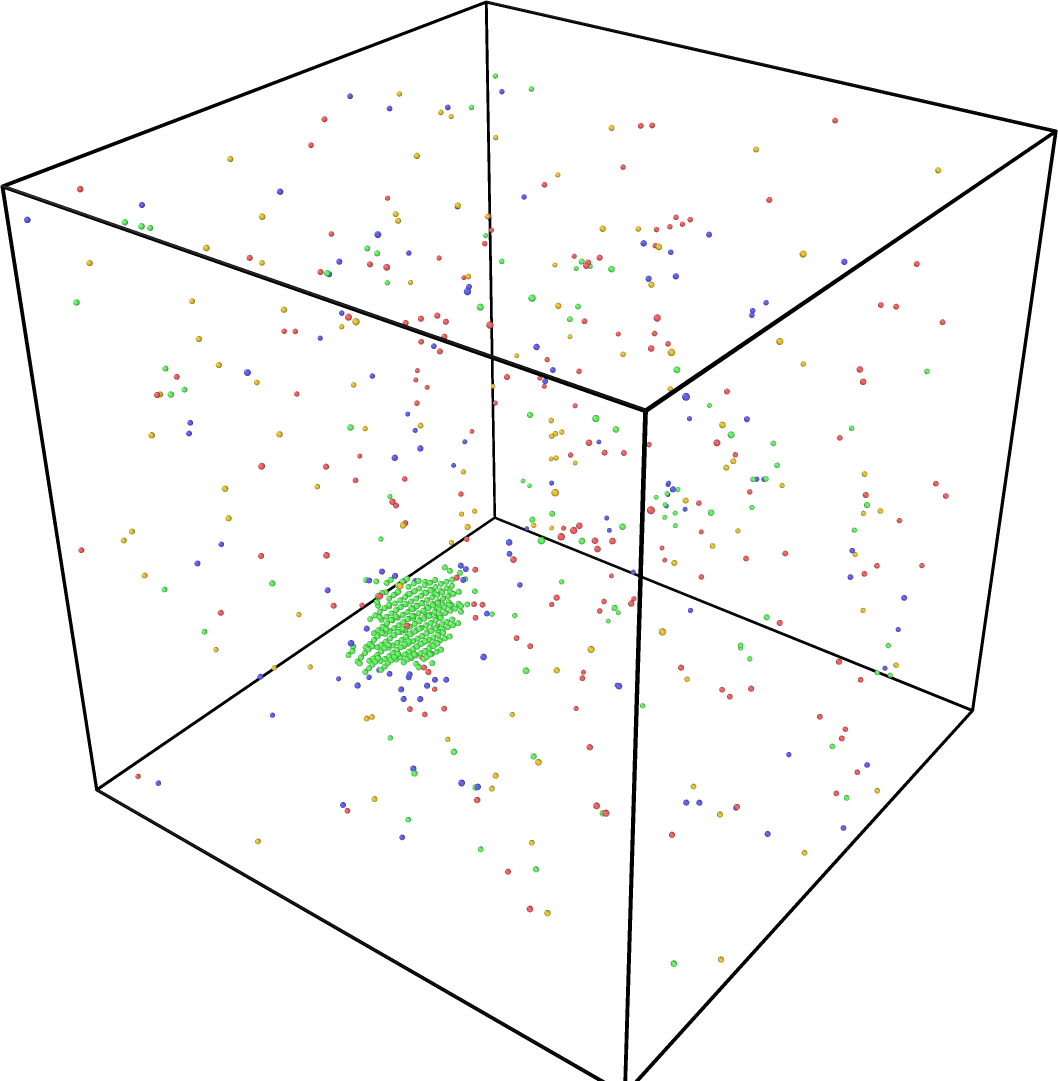}
    \includegraphics[width=0.24\textwidth]{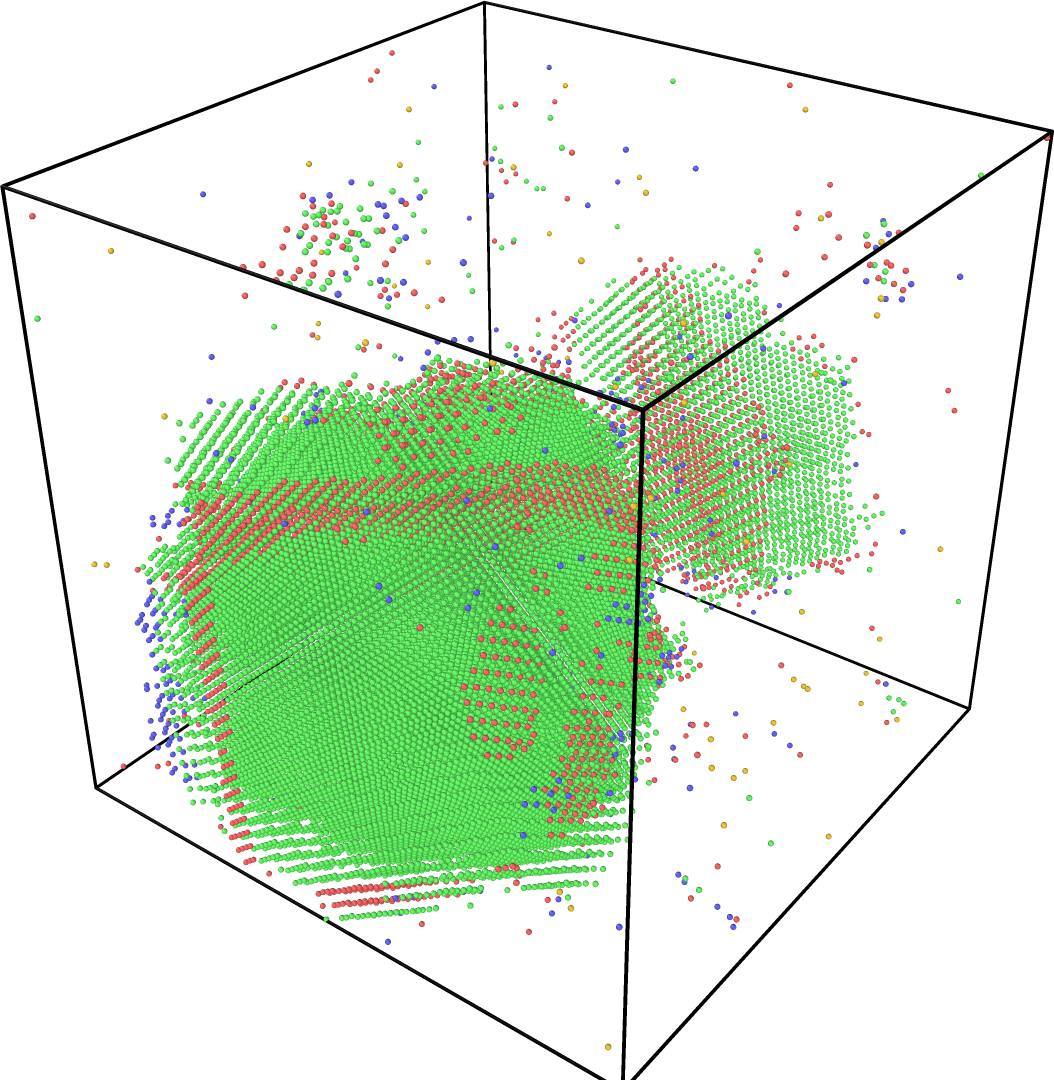}
    \caption{Snapshots of the pure Ni system during nucleation, at $1.76$, $1.77$, $1.79$, and $1.83$ ns. Particles are colored according to their CNA structure, blue for bcc, red for hcp, green for fcc, and unidentified atoms being translucent.} 
    \label{fig:snapshots:100}
\end{figure}

\begin{figure}
    \centering
    \includegraphics[width=0.4\textwidth]{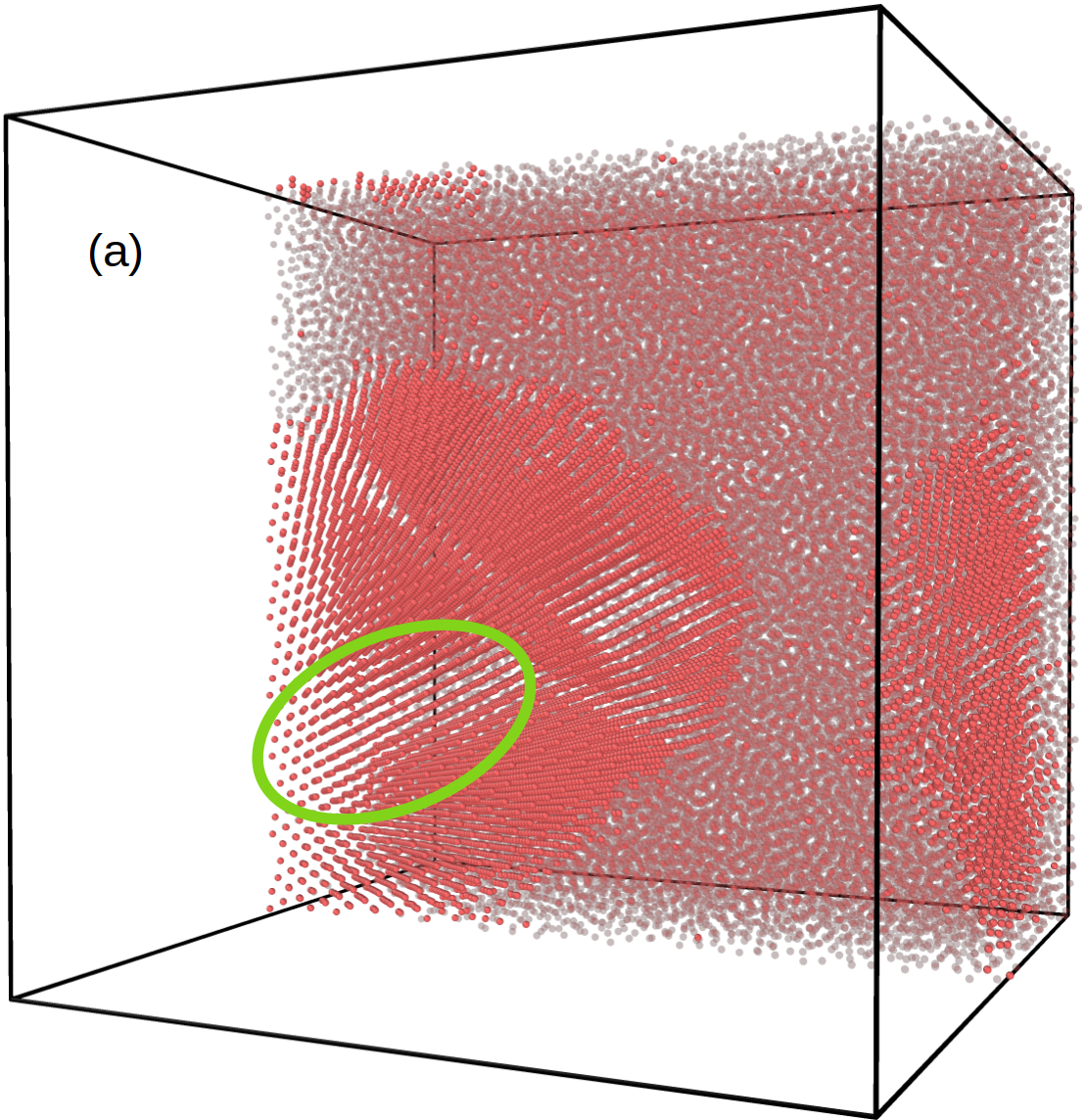}
    \includegraphics[width=0.5\textwidth]{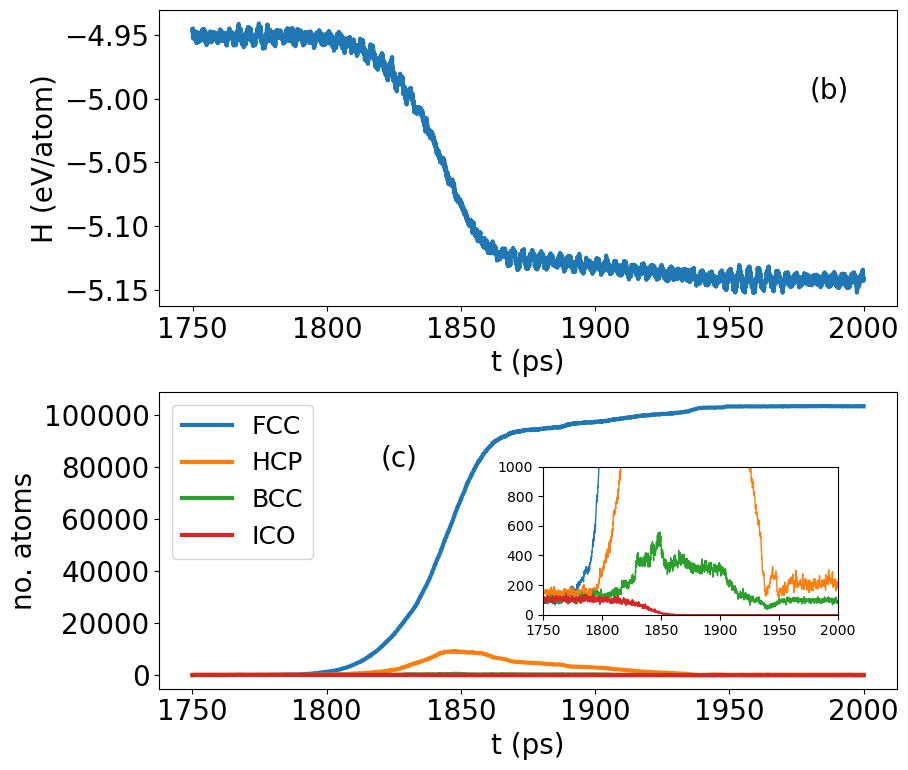}
    \caption{Slice of the simulation box during nucleation at $1.83$ ns, showing the fcc structure of the nucleus  (a). Enthalpy as a function of time during nucleation (b). Abundance of local structures, identified with the CNA, in the pure Ni system (c).} 
    \label{fig:timeseries:100}
\end{figure}

Finally, nucleation pathway of the pure Ni system is investigated. 
The simulation is performed in the NPT ensemble at ambient pressure with a box of the same size as for Al\textsubscript{50}Ni\textsubscript{50}. Nucleation starts around $1.77$ ns after the initial quench from the liquid state at $2000$ K to $1200$ K, as shown in the snapshots in Figure \ref{fig:snapshots:100}. 
A CNA analysis on the inherent structures of the snapshots seems to indicate a single fcc nucleus, emerging directly from the melt through a single step process.
Analysing the bond-order parameters yields an identical picture, as illustrated by the abundance of different structures shown in Figure \ref{fig:timeseries:100}.
Here we observe an increase in fcc structures, with some hcp inclusions corresponding mainly to stacking faults in the fcc crystalline nucleus, that is deemed to disappear in a later stage of the crystallisation.
The single-step process observed here is in contrast to the two-step process reported previously by Orihara \textit{et al} \cite{orihara2020molecular} using an embedded atom model, but is analogous to the single-step nucleation we have observed for pure Al \cite{jakse2022machine} using a HDNNP for that element.

The fact that Al\textsubscript{50}Ni\textsubscript{50} nucleates in a B2 crystalline structure despite the fact that the pure elements counterparts, namely Al and NI, display fcc crystallisation could be understood from the Al-Ni chemical affinity that preexist in the liquid state and is the strongest at equatomic composition, as pointed out in the preceding section. 
Such an affinity favors the compound formation and the corresponding chemical ordering might take place prior to the topological one when the nucleation takes place as was shown in a previous contribution \cite{becker2022crystal}.   

\section{Conclusion}
\label{sec:conclusion}
This work was devoted to the study of homogeneous nucleation of a binary Al-Ni alloy by means of molecular dynamics with a HDNNP trained on AIMD trajectories.
A reasonable agreement is found between our simulation results and experimental data in the liquid state for the local structural properties as well as for the dynamics through self-diffusivity and viscosity. We take this as an indication of the reliability of the HDNNP in describing kinetic processes. The thus validated potential allowed us to perform large scale simulations at an unprecedented accuracy for this alloy system, close to the \textit{ab initio} one.
Simulation of a highly undercooled equimolar melt shows a single step homogeneous nucleation into a B2 phase. 
The resulting supercritical nuclei have an irregular shape, in disagreement with the assumptions of the classical theory, and the critical size is estimated to be between $80$ and $120$ atoms.
In the pure Ni system with the same HDNNP, a single-step nucleation process towards the fcc phase is also seen, in contrast to previous results using EAM.
This, at the very least, highlights the sensitive dependence of the nucleation pathway on the details of the underlying interatomic potential.

While the focus was on just two compositions in the present work, the HDNNP is highly accurate for the melt dynamics also over the entire composition range, and thus the nucleation pathways also of other alloy compositions can be studied. We deem it important that the validation occurs against liquid-state transport coefficients: diffusion and interdiffusion are important processes in nucleation and growth, while empirical potentials are often gauged against the equilibrium crystal structures.

\section*{Acknowledgments}
We acknowledge the CINES and IDRIS under Project No. INP2227/72914, as well as CIMENT/GRICAD for computational resources. We acknowledge financial support under the French-German project PRCI ANR-DFG SOLIMAT (ANR-22-CE92-0079-01). This work has been partially supported by MIAI@Grenoble Alpes (ANR-19-P3IA-0003).
JS acknowledges funding from the German Academic Exchange Service through DLR-DAAD fellowship grant number 509.
We thank Fan Yang for fruitful discussion on the experimental total structure factors.

\bibliographystyle{unsrt}
\bibliography{references}

\end{document}